\documentclass[aps,prc,preprintnumbers,fleqn,showpacs,showkeys,
superscriptaddress,floatfix]{revtex4}
\usepackage{amsmath,amsfonts,bm}
\usepackage[dvipdfm]{graphicx}
\usepackage[dvips]{color}

\begin{document}
\preprint{INHA-NTG-04/2011}
\title{Nucleon properties in nuclear matter}

\author{Ulugbek Yakhshiev }
\email{yakhshiev@inha.ac.kr}

\affiliation{Department of Physics, Inha University, Incheon
402-751, Republic of Korea }

\author{Hyun-Chul Kim}
\email{hckim@inha.ac.kr}

\affiliation{Department of Physics, Inha University, Incheon
402-751, Republic of Korea }

\date{March, 2011}

\begin{abstract}
We present recent studies on the effective mass of the nucleon in 
infinite and homogeneous nuclear matter and its relation to nuclear
matter properties within the framework of the in-medium modified
Skyrme model. Medium modifications are achieved by 
introducing optical potential for pion fields and parametrization of
the Skyrme parameter in nuclear medium. The present 
approach is phenomenologically well justified by pion physics in
nuclear matter and describe successfully bulk nuclear matter
properties.  
\end{abstract}


\maketitle

%%%%%%%%%%%%%%%%%%%%%%%%%%%%%%%%%%%%%%%%%%%%
%% MAINMATTER
%%%%%%%%%%%%%%%%%%%%%%%%%%%%%%%%%%%%%%%%%%%%

\section{Introduction}
The equation of state (EOS) has been one of the most important topics
in nuclear many-body problems. In-medium modified Skyrme
model~\cite{Rakhimov:1996vq} and especially its recently modified
version~\cite{Yakhshiev:2010kf} allow us to investigate not only
single nucleon properties in nuclear matter, but also the matter as a 
whole~\cite{Rakhimov:1996vq,Yakhshiev:2010kf}\footnote{For more
  references devoted to nuclear many-body problem see the references
  presented in the work~\cite{Yakhshiev:2010kf}.}.  In its initial
version, this approach has been used to investigate the properties of
the single nucleon in nuclear matter, pionic
clouds~\cite{Rakhimov:1996vq} being considered in determining external
parameters. As a more realistic approach, we need to modify the core
part of the nucleon, which provides a possibility to reproduce the
binding energy of the system and to discuss the thermodynamic
properties of the bulk matter~\cite{Yakhshiev:2010kf}. 
While the core of the nucleon in the Skyrme model is represented by
the Skyrme term in the Lagrangian, its modification is related to the 
change of the Skyrme parameter. In more detailed treatments,
Skyrme's quartic stabilizing term may 
be revised by explicit vector meson degrees of freedom. In this 
sense, the core modifications of the nucleon may be pertinent to the
vector meson properties in nuclear matter. In the present contribution
we will describe the peculiarities of our approach. 

In Ref.~\cite{Rakhimov:1996vq}, the initial in-medium modified Skyrme
Lagrangian was presented, in which the modifications were achieved by
changing the pion mass in nuclear medium by means of the nonlocal
optical potential. Due to the nonlocality, not only the mass term but
also the kinetic term in the Lagrangian were changed. 
That procedure does not affect the soliton core, while modifications
were exact only in linear approximation and by assumption it was
extrapolated to the nonlinear case. In general, one should take into
account also the modifications of the quartic term. This is especially
required in the case of high density considerations. In this regard,
we need to modify also the Skyrme's stabilizing term, i.e., to change
the Skyrme parameter as a function of the external nuclear density. 
The resulting Lagrangian is given as follows:
\begin{eqnarray}
{\cal L}^*&=&\frac{F_\pi^2}{16}\,{\rm Tr}\left(\frac{\partial
U}{\partial t}\right)\left(\frac{\partial U^\dagger}{\partial
t}\right) -\frac{F_\pi^2}{16}\,\alpha_p({\vec r}){\rm
  Tr}({\vec\nabla} U)\cdot({\vec\nabla}
 U^\dagger)\cr
&&+\frac{1}{32e^2\gamma({\vec r})}\,{\rm
Tr}[U^\dagger\partial_\mu U,U^\dagger\partial_\nu U]^2
+\frac{F_\pi^2m_\pi^2}{16}\,\alpha_s({\vec r}){\rm Tr}(U+ 
U^\dagger-2)\,, \label{Eq:Lag}
\end{eqnarray}
where $F_\pi$ denotes the pion decay constant, $e$ is the Skyrme
parameter, and $m_\pi$ stands for the pion mass. The medium
functionals, $\alpha_s$, $\alpha_p$ and $\gamma$, are written in the
following forms
\begin{eqnarray}
\alpha_s=1-\frac{4\pi b_0\rho({\vec r})f}{m_\pi^{2}},\quad
\alpha_p=1-\frac{4\pi c_0\rho({\vec r})}{f+g_0'4\pi
c_0\rho({\vec r})},\quad \gamma=\exp\left(-\frac{\gamma_{\rm
    num}\rho({\vec r})}{1+\gamma_{\rm 
den}\rho({\vec r})}\right)\,.
\label{medfunc}
\end{eqnarray}
They represent the influence of the surrounding environment on the
properties of the single skyrmion. The parameters $\alpha_s$ and $\alpha_p$
are related to the corresponding phenomenological $S$- and $P$-wave
pion-nucleus scattering lengths and volumes, i.e. $b_0$ and $c_0$,
respectively, and describe the pion physics in a baryon-rich 
environment~\cite{Ericsonbook}. The last functional $\gamma$
represents the modifications of the skyrme parameter with two
variational parameters 
$\gamma_{\rm num}$ and $\gamma_{\rm den}$~\cite{Yakhshiev:2010kf}.
The density of the surrounding
nuclear environment is given by $\rho$, $g_0^{\prime}$ denotes the
Lorentz-Lorenz or correlation parameter, $f=1+m_\pi/m_{N}^{\rm free}$
represents the kinematical factor, and $m_{N}^{\rm free}$ is the
nucleon mass in free space.

The Lagrangian in Eq.~(\ref{Eq:Lag}) will be used to calculate properties
of the nucleons in nuclear matter and the bulk matter properties.
The parameters of the model are fitted to be
$F_\pi=108.78$~MeV and $e=4.85$ so as to reproduce the
experimental values of the nucleon and $\Delta$ in free space.
The pion mass is also fixed to be its experimental value,
$m_\pi=m_{\pi^o}^{\rm exp}=134.98$~MeV. A set of values of parameters
in the medium functionals~(\ref{medfunc}) are taken from the
analysis of phenomenological data for pion-nucleus
scattering~\cite{Ericsonbook}.
Since the environment acting on the single nucleon properties has
a homogeneous and constant density, one can choose the spherically
symmetric ``hedgehog" form for the boson field 
$U=\exp\{i\hat{{\vec n}}\cdot{\vec \tau}F(r)\}$, where
${\vec n}$ denotes the unit vector in coordinate space and
${\vec \tau}$ are the usual Pauli matrices.
Then the problem will be much simplified and thereafter we will follow
this choice. 

\section{Nuclear matter properties \label{sec:Lag}}
\begin{figure}
  \includegraphics[height=.35\textheight]{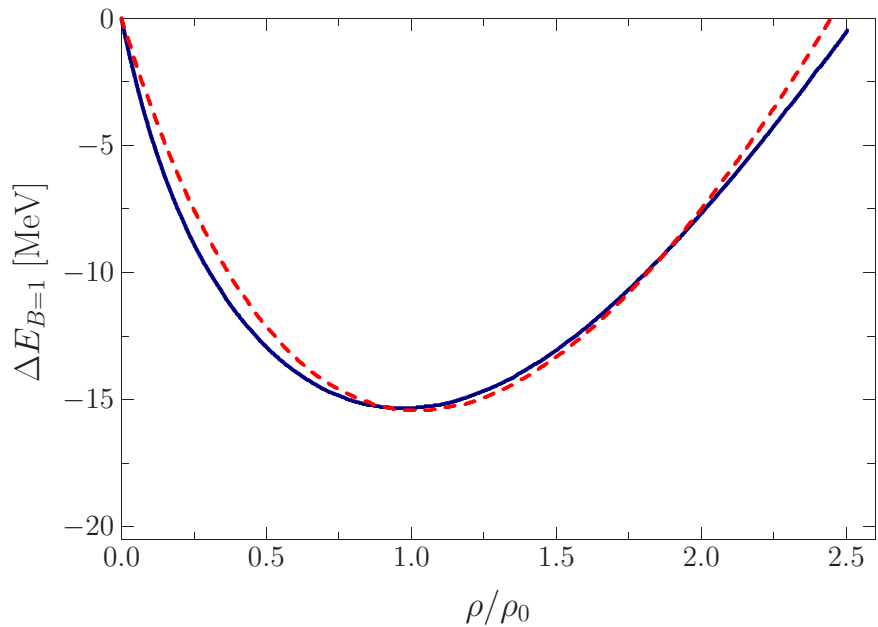}
  \caption{ The binding energy per nucleon as a function of
$\rho/\rho_0$. The solid curve corresponds to the results with
$\gamma_{\rm num}=2.1m_\pi^{-3}$, $\gamma_{\rm den}=1.45m_\pi^{-3}$
and $P$-wave scattering volume $c_0=0.21m_\pi^{-3}$.  
The dashed one draws the case when $\gamma_{\rm
num}=0.8m_\pi^{-3}$, $\gamma_{\rm den}=0.5m_\pi^{-3}$ and $P$-wave
scattering volume $c_0=0.09m_\pi^{-3}$. The $S$-wave scattering length
has the value $b_0=-0.024m_\pi^{-1}$.  The correlation parameter
is fitted near its experimental value $g_0'=0.7$. 
}
\label{bindingEN}
\end{figure}

The results of the binding energy per nucleon are shown in
Fig.~\ref{bindingEN}. 
The solid and dashed curves 
draw the parametrization of $\gamma$ in Eq.~(\ref{medfunc}).
Our minimization procedure has been performed in such a way
that the values of the variational parameters, i.e., $\gamma_{\rm
  num}$ and $\gamma_{\rm den}$, lead to the 
minimum of the binding energy per nucleon, defined crudely as
\begin{equation}
\Delta E_{B=1}=m_{ N}^*(\rho)-m_{ N}^{\rm free}\,,
\label{BE}
\end{equation}
at the normal nuclear matter density.
The difference between the two curves in Fig~\ref{bindingEN} arise from
the different values of the $P$-wave scattering length, i.e. 
$c_0=0.21m_\pi^{-3}$ corresponds to the solid curve and
$c_0=0.09m_\pi^{-3}$ to the dashed one. One can see that the
dependence on the density is rather insensitive to the changes of
$P$-wave scattering volume. The effect of the changes in $b_0$ is even
smaller. 

\begin{figure}
  \includegraphics[height=.35\textheight]{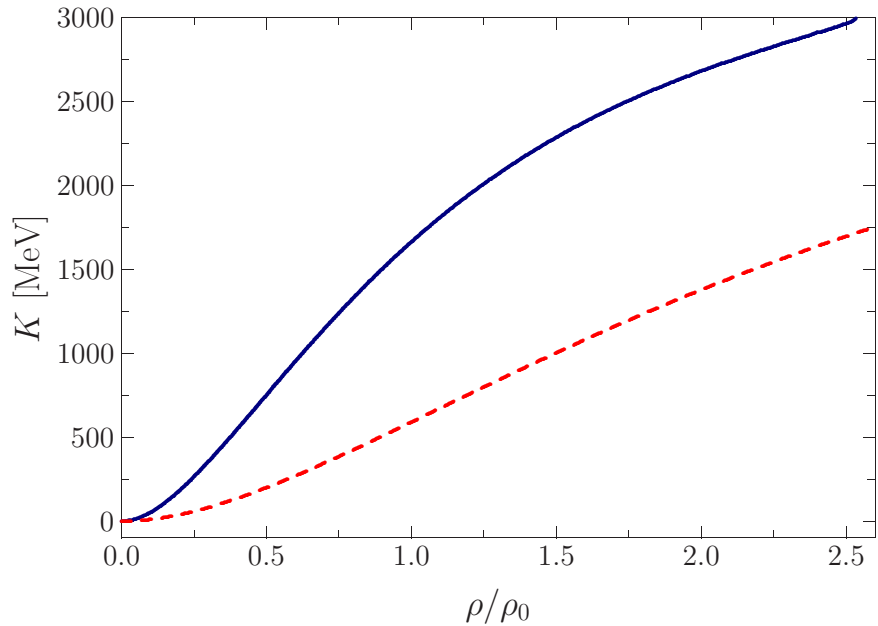}
  \caption{The compressibility as a function of
$\rho/\rho_0$. The other notations are the same as in
Fig.~\ref{bindingEN}. } 
\label{compress}
\end{figure}

However, another important quantity is the compressibility of nuclear
matter defined as 
\begin{equation}
K=9\rho_0^2\left.\frac{\partial^2\Delta E_{B=1}}{\partial
    \rho^2}\right|_{\rho=\rho_0}\,,
\end{equation}
which depends strongly on $c_0$. The corresponding result is presented
in Fig.~\ref{compress}. 
For example, at the empirical value of $c_0=0.21m_\pi^{-3}$, the
compressibility turns out to be very large 
($K\sim 1640$~MeV). The other approaches like relativistic
Dirac-Brueckner-Hartree-Fock~\cite{TerHaar:1986ii,Brockmann:1990cn} 
or the Walecka model~\cite{Serot:1984ey} give much more smaller values
of the compressibility. Lowering the value of $c_0$ leads to the
noticeably 
decreasing value of $K$: if we use $c_0=0.09m_\pi^{-3}$,
the compressibility decreases and becomes consistent with that of the
Walecka model ($K\sim 580$~MeV). Lowering further the value up to
$c_0=0.06m_\pi^{-3}$ give the result $K\sim 300$~MeV that is
comparable with experimental value of $K$ and with that in 
Dirac-Brueckner-Hartree-Fock approaches. Our  
conclusion is that the smaller values of $c_0$ than that used  
in the pionic atom analysis will be preferable as far as the
compressibility is concerned. The compressibility $K$ is sensitive to
the position of the saturation point. If one fits the 
saturation point at slightly lower densities, it is found that the
compressibility drastically decreases. 

One can also discuss the symmetry energy in a crude
approximation~\cite{Klebanov:1985qi,Lee:2003aq} 
\begin{equation}
  \label{eq:11}
E_{\rm sym}=\frac{1}{12}\,m_{\Delta- N}^*\,,
\end{equation}
where $m_{\Delta- N}^*$ is the effective $\Delta-N$ mass difference. 
We obtained the following results for the symmetry energy:
$E_{\rm sym}(\rho_0)\approx 14.19$~MeV for
$c_0=0.09m_\pi^{-3}$ whereas $E_{\rm  sym}(\rho_0)\approx 8.71$~MeV
for $c_0=0.21m_\pi^{-3}$. These results slightly are underestimated in
comparison with its  
experimental value $E_{\rm sym}\sim
20 - 30$~MeV. Further generalizations of the model including the
isospin breaking effects, finite nuclei corrections may improve the
situation.

\section*{Acknowledgments}
Authors would like thank organizers of the Baryon 2010 for the
hospitality during their stay at Osaka University. 
The present work is
supported by Basic Science Research Program through the National
Research Foundation of Korea (NRF) funded by the Ministry of
Education, Science and Technology (grant number: 2009-0089525).

\end{document}